\documentclass[aip,jap,showpacs,reprint]{revtex4-1} %reprint,twocolumn,groupedaddress
\usepackage{amsmath} % AMS Math Package
\usepackage{amsthm} % Theorem Formatting
\usepackage{amssymb}	% Math symbols such as \mathbb
\usepackage{graphicx} % Allows for eps images
\usepackage[usenames,dvipsnames]{pstricks}
\usepackage{pst-grad} % For gradients
\usepackage{pst-plot} % For axes
\usepackage{color}
\usepackage{wasysym}
\usepackage{url}
\usepackage{bbold}
\usepackage{bm}
\newcommand{\markup}[1]{{\color{black}{#1}}}
\newcommand{\highlight}[1]{{\color{black}{#1}}}
\newcommand{\high}[1]{{\color{black}{#1}}}
\newcommand{\highl}[1]{{\color{black}{#1}}}
\newcommand{\highli}[1]{{\color{black}{#1}}} 
\newcommand{\highlig}[1]{{\color{black}{#1}}}
\usepackage{subfigure}
\usepackage{gnuplottex}

\renewcommand{\v}[1]{\ensuremath{\mathbf{#1}}} % for vectors
\newcommand{\avg}[1]{\left< #1 \right>} % for average
\let\baraccent=\= % rename builtin command \= to \baraccent
\renewcommand{\=}[1]{\stackrel{#1}{=}} % for putting numbers above =

\theoremstyle{definition}

\theoremstyle{remark}


\begin{document}
% Use the \preprint command to place your local institutional report
% number in the upper righthand corner of the title page in preprint mode.
% Multiple \preprint commands are allowed.
% Use the 'preprintnumbers' class option to override journal defaults
% to display numbers if necessary
%\preprint{}

\title{Influence of interface potential on the effective mass in Ge nanostructures}

\author{E. G. Barbagiovanni}
\email[]{eric.barbagiovanni@ct.infn.it}
\affiliation{MATIS IMM-CNR and Dipartimento di Fisica e Astronomia, Universit\`{a} di Catania, via S. Sofia 64, 95123 Catania, Italy}

\author{S. Cosentino}
\affiliation{MATIS IMM-CNR and Dipartimento di Fisica e Astronomia, Universit\`{a} di Catania, via S. Sofia 64, 95123 Catania, Italy}

\author{D. J. Lockwood}
\affiliation{Measurement Science and Standards, National Research Council, Ottawa, Ontario K1A 0R6, Canada}

\author{R. N. Costa Filho}
\affiliation{Departamento de F\'{i}sica, Universidade Federal do Cear\'{a}, Caixa Postal 6030, Campus do Pici, 60455-760 Fortaleza, Cear\'{a}, Brazil}

\author{A. Terrasi}
\affiliation{MATIS IMM-CNR and Dipartimento di Fisica e Astronomia, Universit\`{a} di Catania, via S. Sofia 64, 95123 Catania, Italy}

\author{S. Mirabella}
\affiliation{MATIS IMM-CNR and Dipartimento di Fisica e Astronomia, Universit\`{a} di Catania, via S. Sofia 64, 95123 Catania, Italy}

\date{\today}

\begin{abstract}

The role of the interface potential on the effective mass of charge carriers is elucidated in this work. We develop a new theoretical formalism using a spatially dependent effective mass that is related to the magnitude of the interface potential. Using this formalism we studied Ge quantum dots (QDs) formed by plasma enhanced chemical vapour deposition (PECVD) and co-sputtering (sputter). These samples allowed us to isolate important consequences arising from differences in the interface potential. We found that for a higher interface potential, as in the case of PECVD QDs, there is a larger reduction in the effective mass, which increases the confinement energy with respect to the sputter sample. We further understood the action of O interface states by comparing our results with Ge QDs grown by molecular beam epitaxy. It is found that the O states can suppress the influence of the interface potential. From our theoretical formalism we determine the length scale over which the interface potential influences the effective mass.

\end{abstract}


\maketitle

\section{Introduction \label{intro}}

Si and Ge nanostructures (NSs) remain \highlight{amongst} the most important materials for photovoltaic and electronic applications \cite{Ray:2013, Mangolini:2013}. Si is extensively used in \highlight{the microelectronics} industry partly due to the importance \high{and stability} of its oxide \cite{Zwanenburg:2013}. \highlight{The Ge oxide states can create complications in NSs.} Fabrication of Ge NSs in a SiO$_2$ matrix \highlight{often results in} a high concentration of interface defect states \cite{Min:1996, Barbagiovanni:2013, Barbagiovanni:2012}, because Ge can easily \highlight{occupy} the Si position within the SiO$_2$ matrix due to its low formation energy. Sub-oxide interface states give rise to a defect photoluminescence (PL) band that can dominate quantum confinement (QC) effects \cite{Barbagiovanni:2013}. However, with careful control over the fabrication \cite{Lockwood:2013} and characterization \cite{Cosentino:2014} conditions, Ge NSs do clearly demonstrate QC effects. Furthermore, because of a larger Bohr radius in Ge compared to Si, Ge \highlight{has the advantage of greater tunability} in the gap energy ($E_G$) \highlight{due to QC} \cite{Cosentino:2014_1}. Moreover, it was recently demonstrated that interface defect states may enhance the efficiency of photovoltaic devices due to preferential hole trapping \cite{Cosentino:2014_1}.

The nature of \highlight{its} \highlight{interfacial structure} is arguably the defining parameter of a NS. Fundamentally, the introduction of an interface modifies bulk parameters \cite{Barbagiovanni:2013} and introduces new phenomena. Interface states can break valley degeneracy \cite{Dusko:2014}. Dangling bonds, stress, and defect states elicit the degree to which the interface potential confines charge carriers \cite{Barbagiovanni:2013}. Hydrogen versus oxygen termination at the interface influences the oscillator strength \cite{Guerra:2010} and creates polarization effects at the interface \cite{Guerra:2011}. Nanostructure sensing devices rely on the charge configuration at the interface, which `communicates' with the environment \cite{Mattioli:2014} and determines the conductivity within the NS \cite{Ruhle:2012}. Nonetheless, the role of the interface remains largely misunderstood \cite{Dohnalova:2014}. The exact band energy alignment at the interface remains a topic of intense research \cite{Seguini:2013}, as it is difficult to establish the exact chemical environment at the interface \cite{Hirose:2007}. Defect engineering at the interface can have a complicated effect on the QC of the charge carriers \cite{Zacharias:2012, Baldovino:2011}, which can be difficult to model theoretically \cite{Bagolini:2010, Lusk:2014}.

The crystal potential changes across the interface of a NS and thus the \highlight{carrier} effective mass (EM) changes\highlight{, which is modelled} by invoking the Bastard-type boundary conditions \cite{Bastard:1981, Chetouani:1995, Ganguly:2006, Moskalenko:2007}. There is experimental \cite{Barbagiovanni:2012, Shannon:1993, Cosentino:2013, Seas:1997} and theoretical \cite{Seino:2011, Barbagiovanni:2013_1} evidence that the EM should depend on \highlight{the} NS dimension. Understanding how the EM varies in low-dimension remains a challenge as experimental measurements are model dependent \cite{Lockwood:1996, Rossner:2003} and it is difficult to theoretically scale the EM to low dimensions \cite{Barbagiovanni:2012, Seino:2012}. Density functional theory (DFT) was used to calculate the EM from the density of states \cite{Seino:2011}. However, it is not clear how to model the interface nor the excited states within DFT \cite{Barbagiovanni:2011, Moskalenko:2007}. 

A common pitfall in theoretical modelling is the lack of information regarding the interface atomic potentials, which leads to extensive use of H-terminated surfaces at the \highlight{expense} of the experimental details \cite{Barbagiovanni:2013}. \highl{Sub-oxide interfaces are typically modelled using OH termination \cite{Guerra:2013, Konig:2008}. The local density approximation is used to correct for the local charge environment to obtain agreement with the experimental optical gap \cite{Guerra:2013, Barbagiovanni:2013}. \highli{However, there are recent theoretical results using embedding matrices, such as SiC \cite{Weissker:2002_1}. Ref. \onlinecite{Seino:2012} embedded Si NSs in a SiO$_2$ matrix by inserting O atoms between Si-Si bonds followed by a randomizing procedure leading to defect and stress free structures.} Therefore, relaxed assumptions are required to obtain agreement with experiment and the defect landscape cannot be readily included \cite{Broqvist:2009, Seino:2010}.}

Owing to the rich variety of phenomena that the interface can give rise to, understanding the contribution of the individual components can be difficult. \highlight{Indeed,} theoretical studies have suggested\cite{Niquet:2000, Barbagiovanni:2012, Barbagiovanni:2013} $E_G\sim D^{-x}$, where $1\leq x\leq 2$ and $D$ is the NS diameter, which has been demonstrated experimentally \cite{Mirabella:2013, Kovalev:1998, Barbagiovanni:2013}. In this work, we elucidate the \highlight{precise} role of the interface potential on the EM. An analytic formalism for the EM and interface \cite{Barbagiovanni:2013_1} was developed and applied to study Ge quantum dots (QDs) grown by molecular beam epitaxy (MBE) \cite{Barbagiovanni:2013_3}. The parameters developed during our study on MBE Ge QDs are transferred to the present study on rf-magnetron co-sputtering (sputtering), and plasma enhanced chemical vapour deposition (PECVD) Ge QDs. \highlight{The PECVD and sputter samples were chosen for this work because they were fabricated under similar experimental conditions, \high{but exhibit} a measurable variation in the chemical structure of the interface.} Both sputtered and PECVD \highlight{QDs} have the same dimensional dependence as the MBE QDs, which allows us to isolate the role of the interface potential on the EM. 

\highlight {Comparing PECVD and sputter samples,} it \highlight{is} found that as the interface potential increases the EM is reduced thus increasing the confinement energy. We determine the length scale at which the EM is modified due to the interface potential. Furthermore, \highlight{by comparing with MBE Ge QDs} we find that O interface states can suppress this mechanism.  These results are important for solar cell applications, as we recently demonstrated that defects at the interface can \highlight{be engineered to enhance the light harvesting capabilities in Ge QDs} \cite{Cosentino:2014_1}.

\section{Experiment \label{expt}}

\markup{In this work, we focus our analysis on experimental results from two fabrication methods for Ge quantum dots (QDs) that produce different interface potentials.} Thin films of Ge-rich SiO$_2$ were deposited using either PECVD \cite{Cosentino:2014} or sputtering \cite{Cosentino:2014_1}. Thermal annealing was carried out at 800 $^o$C or 600 $^o$C in N$_2$ for PECVD or sputtered samples, respectively, to promote Ge QD nucleation. \highlight{In Refs. \onlinecite{Cosentino:2014, Cosentino:2014_1}, the experimental details of diameter control and determination are discussed along with crystallinity measurements.} Both PECVD and sputtered QDs were determined to be in an amorphous state with a bulk gap energy ($E_G(\infty)$) of 0.8 eV. \markup{The average diameter in the sputter samples varies from 2 to 4 nm, while in PECVD samples it varies from 3.5 to 8.4 nm. Optical absorption measurements were performed to assess the variation in $E_G$ with QD diameter.} \highli{Ge NSs embedded in an oxide matrix typically demonstrate a PL spectrum dominated by a high concentration of sub-oxide related defects at the interface between the Ge NS and the oxide matrix, thus masking the QC related PL band \cite{Barbagiovanni:2013}.} \highl{In Refs. \onlinecite{Cosentino:2014} and \onlinecite{Cosentino:2014_1}, the absorption spectrum was considered over the PL spectrum, because defects states in the GeO$_2$ interface made it difficult to measure the variation of $E_G$ with QD diameter from the emission spectrum \cite{Cosentino:2011}, \highlig{which is more sensitive to mid-gap defects \cite{Zacharias:1997}}.} \highli{Finally, it is worth noting that due to NS size distribution and carrier leakage there is some inherent error in associating the NS experimental $E_G(D)$ with the correct experimental $D$, \highlig{see Ref. \onlinecite{Barbagiovanni:2013} for more details.}}

\highl{From the absorption spectra, the oscillator strength, and $E_G(D)$ was found in Refs. \onlinecite{Cosentino:2014, Cosentino:2014_1} for each $D$ using the Tauc method. The applicability of the Tauc method to NSs is discussed in detail in Ref. \onlinecite{Cosentino:2014_1}. The Tauc method is based on a semi-empirical model for amorphous systems, which assumes parabolic band edges and optical inter-band transitions between quasi localized states (i.e. momentum is not conserved) \cite{Tauc:1974}. Indeed, the method assumes that the dipole-matrix element (proportional to the oscillator strength) is not $\v{k}$-dependent. Breaking of momentum ($\v{k}$-vector) conservation is a fundamental tenant of quantum confinement in NSs. Furthermore, our NSs are amorphous and thus lack long range order. In addition, the parabolic band edge picture can still be applied in a NS as discussed in the next section. A deviation from linearity happens at low $\sigma h\nu$ due to Urbach tail regions. For this reason, the estimation of $E_G$ for amorphous materials is usually performed for values of $\sigma$ larger than 2$\times$10$^{-9}$ cm$^2$ (for our samples) where a clear linear trend in the Tauc plots appears \cite{Knief:1999}. \highlig{Furthermore, the $E_G$ was calculated from scanning tunnelling spectroscopy measurements on Ge QDs to be $\sim$1.4 eV \cite{Nakamura:2005}. This value agrees with our $E_G$ determined via the Tauc method for a similar diameter range.} Therefore, we find that the Tauc method works well for amorphous NSs and has been experimentally demonstrated in the literature \cite{Knief:1999, Cosentino:2013}.}

\section{Theory \label{theory}}

The analysis of the experimental data is carried out using a variety of theoretical tools. \highl{The approach taken in this work considers an interface-related correction to the EMA. The validity of the EMA has been widely debated in the literature \cite{Barbagiovanni:2013}. Generally, it is assumed that the parabolic band approximation breaks down at low-dimension and thus the EMA cannot be applied. However, recently it has been shown that a notion of a `fuzzy-band-structure' holds for NSs down to $\approx$ 2 nm \cite{Hapala:2013}, and that the $\v{k}\cdot \v{p}$ method does produce the correct symmetry of a QD \cite{Tomic:2011}. The EMA is a first order approximation of the $\v{k}\cdot \v{p}$ generalization that considers transitions only at the conduction band minimum (CBM) and valence band maximum (VBM). In a NS, transitions mainly occur near the Brillouin zone center (the $\Gamma$-point), due to breaking of the momentum selection rules \cite{Barbagiovanni:2012}. \highli{This type of transition is referred to as pseudodirect \cite{Averboukh:2002, Lockwood:2009}. From a theoretical perspective, pseudodirect behavior allows one to ignore phonon events in the formalism.} When a 30-band $\v{k}\cdot \v{p}$ model is considered, spurious solutions are found, which can be removed computationally \cite{Cukaric:2013}. A two-conduction-band $\v{k}\cdot \v{p}$ approach has also been used \cite{Michelini:2011} to describe conduction band hybridization, and a two-band approximation \cite{Barbagiovanni:2012} demonstrates good agreement with experiment \cite{Moskalenko:2007}. A justification of the EMA is given in detail in Refs. \onlinecite{Barbagiovanni:2012, Barbagiovanni:2013}. DFT models including interface corrections have been employed in the literature \cite{Konig:2008, Lusk:2014}, and produce a similar dimensional dependence as the method presented here \cite{Guerra:2013}. However, we opt for the present approach because of the clarity in the physics at the interface. Also, the DFT approaches use a different interface passivation from the actual experiment one.}

First, we \markup{employ} a new theoretical formalism using a spatially dependent effective mass (SPDEM), \markup{as detailed in Ref. \onlinecite{Barbagiovanni:2013_1}} and \highlight{formulated in Refs. \onlinecite{CostaFilho:2011, CostaFilho:2013}}. The important feature here is the introduction of an inverse characteristic length scale: $\gamma^2=2D^{-2}$, where $D$ is the QD diameter \markup{in nanometres}. \highl{$\gamma$ is not tunable, but was derived in Ref. \onlinecite{Barbagiovanni:2013_1} and modifies the dispersion relation of the charge carriers as described in Sec. \ref{disc}.} $\gamma$ acts as a coupling parameter between the momentum of the charge carriers and the confinement potential. In this formalism, the confinement potential, $V_C$, was given by:
\begin{equation}\label{confine_pot}
V_{C,i}=-V_o\exp\left(\frac{2}{D^2\gamma_i^2}\ln^2(1+\gamma_i r_i)\right);
\end{equation}
where \markup{$r_i$ is the particle position in the} $i=x,y,z$ \markup{direction}, and $V_o$ is defined as the energy difference between the QD and the matrix material at the CBM or the VBM for an electron ($V_{o,e}$) or hole ($V_{o,h}$), respectively. Using a point canonical transformation (PCT), we demonstrated that the SPDEM is \markup{directly} related to the confinement potential. This effect is analogous to the relationship between the crystal potential and the bare electron mass \cite{Barbagiovanni:2013_1}. Therefore, \markup{the SPDEM formalism is well adapted to describe the effect of $V_C$ on the effective mass} and yields the dispersion relation: 
\begin{equation}\label{SPDEM}
E_G(D)=E_G(\infty)+
\frac{3\hbar}{\sqrt{2}D}\left[\sqrt{\frac{V_{o,e}}{m_{o,e}^*}}+\sqrt{\frac{V_{o,h}}{m_{o,h}^*}}\right];
\end{equation}
where $m_{o,e}^*$, and $m_{o,h}^*$ are the bulk electron and hole effective mass, respectively. 

The second model used in our analysis is the effective mass approximation (EMA), as detailed in Ref. \onlinecite{Barbagiovanni:2012}. This model describes electron-hole confinement conditions \markup{within an infinite confinement potential context}, according to the relation:
\begin{equation}\label{EMA}
E_G(D)=E_G(\infty)+\frac{7.88}{D^{2}}\;\text{eV}\cdot\text{nm}^2.
\end{equation}
We note that EMA presented here is used to describe \markup{only} the effect of reduced dimension on the electron and hole and includes no additional effects. However, we \markup{discovered a physically significant} phenomenological correction \cite{Barbagiovanni:2013_3} to both the SPDEM and EMA models, which was \markup{applied} to obtain agreement between the two theories, given by:
\begin{eqnarray}\label{MDD}
&\tilde{\mu}(D)=\mu_o e_{SPDEM} D\left(1+\frac{1}{aD^2+bD+c}\right); \label{SP_mu}\\
&\mu(D)=\mu_o e_{EMA} \left(1+\frac{1}{aD^2+bD+c}\right); \label{EM_mu}
\end{eqnarray} 
where a=0.047 nm$^{-2}$, b=0.160 nm$^{-1}$, and c=-0.035 are identical for \markup{both renormalized masses}, $\tilde{\mu}(D)$ and $\mu(D)$, \markup{and were determined in Ref.} \onlinecite{Barbagiovanni:2013_3}. \highl{The parameters `a', `b', and `c' were determined from PL measurements on MBE Ge QDs \cite{Barbagiovanni:2013_3}. Ref. \onlinecite{Rowell:2009} noted that these QDs experience strong exciton localization and thus there is no Stokes shift between the absorption and emission spectrum. According to the theory of QC, localization in position space breaks the momentum selection rules and thus phonon coupling is not required for indirect gap materials \cite{Barbagiovanni:2013}. In Refs. \onlinecite{Barbagiovanni:2013, Barbagiovanni:2012} we noted that a measured Stokes shift was a consequence of interfacial phononic modes (e.g. vibrons). \highli{Strong breaking of the momentum selection rules (or time-reversibility) denoted as `strong confinement' is included in both the EMA \cite{Barbagiovanni:2012} and SPDEM \cite{Barbagiovanni:2013_1} models. We showed previously \cite{Barbagiovanni:2012} that amorphous NSs experience strong confinement in the electron and hole states.}

In our model, the interface potential influences the effective mass, as noted above and discussed in Sec. \ref{disc}.} The parameters $e_{SPDEM}$ \high{in} $\highlight{[nm^{-1}]}$ and $e_{EMA}$ renormalize the bulk reduced mass, $\mu_o$, and will be discussed in Sec. \ref{result}. \highl{Therefore, we conclude that the parameters `a', `b', and `c' are valid for both absorption and emission spectrum, while the influence of the interface is contained in the parameter `e'.

The parameters `a', `b', and `c' are the same for all Ge QD samples and their physical significance is discussed further in Sec. \ref{disc}. Our aim in this work is to understand the influence of the interface on the EM. This influence is contained solely in the parameter `e', which is introduced as correction to the EMA and SPDEM models. This parameter gives us access to the macroscopic features of the interface that would otherwise be inaccessible within the standard EMA. Therefore, when we use Eqs. \eqref{SP_mu} and \eqref{EM_mu} with Eqs. \eqref{SPDEM} and \eqref{EMA}, respectively, we obtain perfect agreement in the dispersion relations \cite{Barbagiovanni:2013_3} allowing us to isolate essential differences in the physics between the two models, as described Sec. \ref{disc}.} \highli{It is important to note that we are presenting a phenomenological analysis and not presenting these models as an alternative to previously existing theoretical models. Our intention is to compare the results of two models across different fabrication methods to extract physically relevant information correlated with the variation in our parameter set.}

\subsection{Methodology \label{method}}

For the sputter and PECVD Ge QDs, we use \markup{Eqs. \eqref{SP_mu} and \eqref{EM_mu}} in the SPDEM-$\tilde{\mu}(D)$ and EMA-$\mu(D)$ (see Ref. \onlinecite{Barbagiovanni:2013_3}) models to determine the relationship between the interface potential and EM. For all QD samples the term $\left(1+\left (aD^2+bD+c\right)^{-1}\right)$ in \markup{Eqs. \eqref{SP_mu} and \eqref{EM_mu}} is the same, thus ensuring that all samples have the same dimensional dependence as found in Ref. \onlinecite{Barbagiovanni:2013_3}. Therefore, we are left to determine only $V_{o,e}$, $V_{o,h}$, and the parameter $e$.

\begin{figure}
\includegraphics[scale=.04]{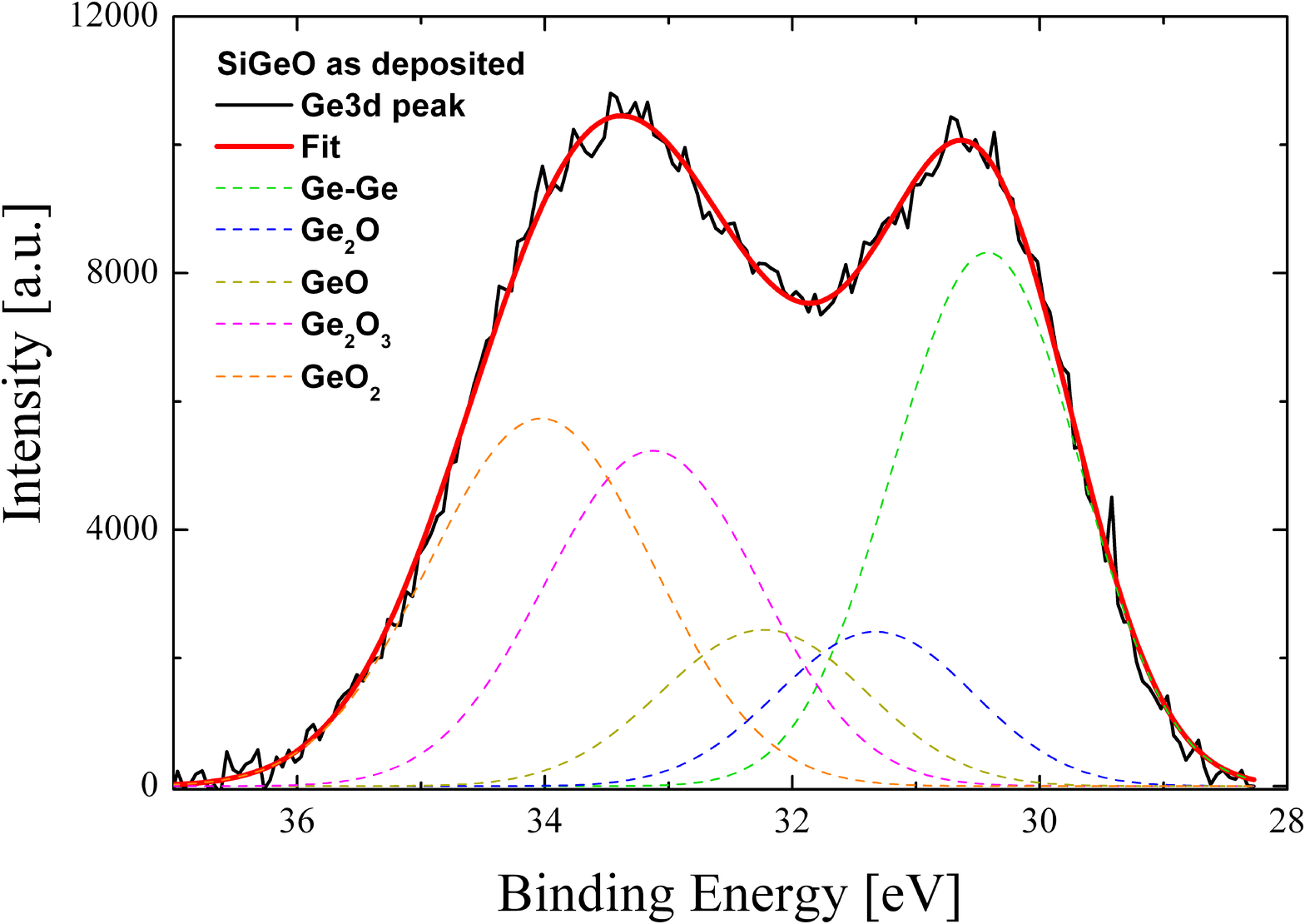}
\caption{XPS measurement of the Ge 3d peak from the as-deposited Ge-rich layer (SiGeO) formed by sputtering. The relative intensity of the Ge-Ge peak with respect to its oxide states is shown by the respective fits. Color online.\label{XPS_Sputter}}
\end{figure}

\begin{figure}
\includegraphics[scale=.85]{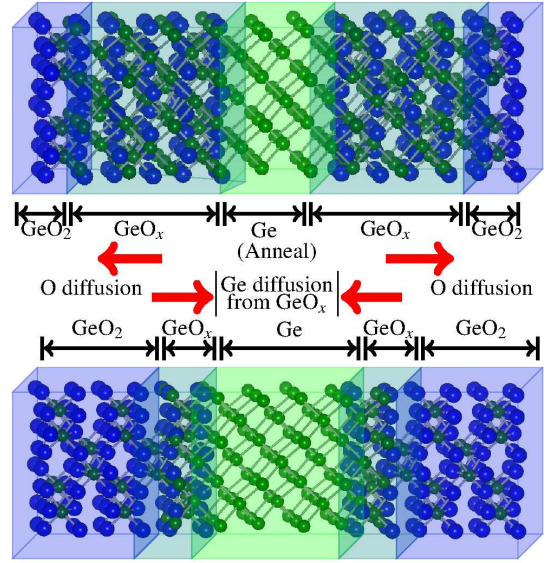}
\caption{Schematic representation of the ideal random mixture model. The top figure depicts the as deposited \markup{Ge-rich layer} where Ge QDs are seen before annealing. Ge QDs are surrounded by a sub-oxide shell and pure GeO$_2$. After annealing, \markup{the} Ge QD diameter \highlight{increases} as stoichiometric GeO$_2$ forms and the GeO$_x$ layer \markup{thickness} decreases. \highlight{The diffusion process is indicated by the arrows in the middle part of the figure.} Molecular structures were generated using VESTA \cite{Momma:2011:Vesta} \markup{and represent only a cross section of the QD structure}. Color online.\label{RMM_fig}}
\end{figure}

First, we consider $V_{o,e}$ and $V_{o,h}$. From x-ray photoemission spectroscopy (XPS) in sputtered Ge QDs, a Ge-Ge signal was measured in the as-deposited film, see Fig. \ref{XPS_Sputter}. This signal indicated that Ge QDs formed upon initial film deposition. \markup{From the XPS results, no Ge-Si nor Si-Si signal} are observed indicating that all of the Ge is either within the QD or in an oxide state, \markup{and all of the Si is in the SiO$_2$ state \cite{Cosentino:2014_1}}. Therefore, sputter thin film deposition, under our given experimental conditions, can be modelled \markup{approximately} by the random mixture model (RMM) \cite{Franzo:2008, Cosentino:2011}. \markup{This model supposes that the film is described by the formation of a mixture of Ge, GeO$_x$, and GeO$_2$ centres \cite{Temkin:1975}}. A representation of the RMM is given in Fig. \ref{RMM_fig}. In the top part of Fig. \ref{RMM_fig}, we indicate a region of Ge surrounded by layer of GeO$_x$ and a thin layer of GeO$_2$. Upon annealing, the \highlight{Ge QD diameter and the stoichiometric oxide concentration increase}. \highlight{This} is a result of Ge diffusing \highlight{from} the \highlight{Ge-rich} GeO$_x$ region \highlight{into the Ge QD region} \cite{Cosentino:2013_2}, which is indicated in the bottom part of Fig. \ref{RMM_fig}. \highlight{Simultaneously, it is possible that O diffuses toward the GeO$_2$ region}. This observation implies that the interface between the Ge QD and the matrix material is comprised of a large concentration of GeO$_x$ states. As a first approximation, we consider a Ge-GeO interface with a potential of approximately $V_{0,e}$ = 0.6 eV and $V_{0,h}$ = 1.8 eV \cite{Binder:2011}.

\begin{figure}
\includegraphics[scale=.85]{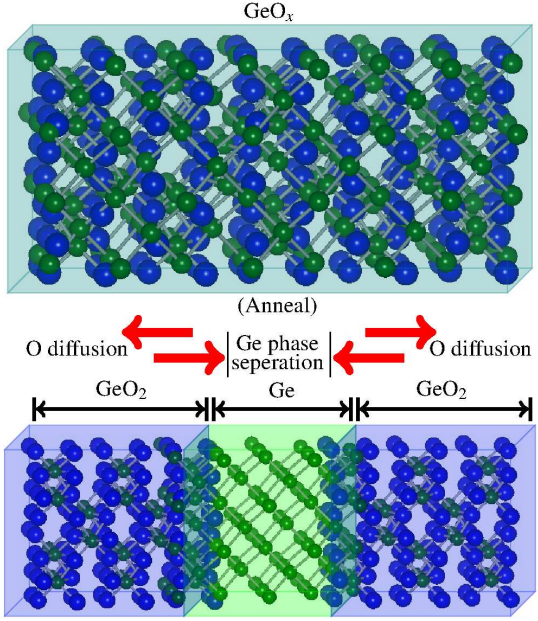}
\caption{Schematic representation of the ideal random bonding model. The top figure depicts the as deposited \markup{Ge-rich layer} comprised of sub-stoichiometric GeO$_2$, GeO$_x$, where $x<$1.5. After annealing, phase separation occurs \markup{and} Ge QDs form surrounded \markup{by} GeO$_2$, with a small concentration of GeO$_x$ right at the interface. \highlight{The diffusion process is indicated by the arrows in the middle part of the figure.} Molecular structures were generated using VESTA \cite{Momma:2011:Vesta} \markup{and represent only a cross section of the QD structure}. Color online.\label{RBM_fig}}
\end{figure}

In the case of PECVD QDs, a different behaviour was observed. Here the QDs are more accurately \markup{modelled} by the random bonding model (RBM), under our given experimental conditions \cite{Franzo:2008}, see Fig. \ref{RBM_fig}. \markup{In the RBM all of the Ge is assumed to be in one of its oxide states \cite{ Philipp:1972}.} \markup{Raman measurements (see Ref. \onlinecite{Cosentino:2014}) indicate that upon deposition the film is primarily in a Ge sub-oxide state, i.e. there is no clear observation of a Ge-Ge signal, as represented in the top part of Fig. \ref{RBM_fig}.} After annealing, GeO$_x$ will phase separate into Ge QDs and GeO$_2$ \highlight{mediated by Ge diffusion}, as represented in the bottom part of Fig. \ref{RBM_fig}. \highlight{As in the case of sputter QDs, it is possible that O undergoes diffusion as well.} This behaviour implies that the interface is mainly composed of GeO$_2$. \markup{Though there must \highlight{still} exist a small concentration \markup{(\highlight{as} compared to sputtered Ge QDs)} of GeO$_x$ \highlight{right} at the interface, which will lower the interface potential \highlight{slightly} from the ideal GeO$_2$ value.} Therefore, we assume that the interface potential is close to the value of Ge-GeO$_2$, which is $V_{0,e}$ = 1.2 eV and $V_{0,h}$ = 3.6 eV \cite{Kobayashi:2009}.

The final step is to determine $e$ for both the SPDEM and EMA models. We consider a Tauc plot of the experimental absorption cross section, $\sigma$, for PECVD and sputtered Ge QDs from Refs. \onlinecite{Cosentino:2014} and \onlinecite{Cosentino:2014_1}, respectively. $\sigma$ is modelled via the relation:
\begin{equation}\label{tauc}
\sigma=\frac{B^*}{h\nu}\left(h\nu-E_G(D)\right)^2;
\end{equation}
where $B^*$ \highlight{is proportional to the} the oscillator strength and was determined experimentally, and $h\nu$ is the photon energy. $E_G(D)$ is given by either \highlight{SPDEM-$\tilde{\mu}(D)$ or EMA-$\mu(D)$,} in Eq. \eqref{tauc}. \highl{$E_G(D)$ and $D$ was determined experimentally in Refs. \onlinecite{Cosentino:2014} and \onlinecite{Cosentino:2014_1}, see Sec. \ref{expt}.} In the case of EMA-$\mu(D)$, we fit Eq. \eqref{tauc} via $e_{EMA}$ to the experimental data for each QD $D$ \highl{and $E_G(D)$}. \markup{For SPDEM-$\tilde{\mu}(D)$, as we noted above, the interface potential was approximated from our discussion of the RMM, or RBM for the sputter, or PECVD samples, respectively.} \markup{Since these models are only a first approximation} we allowed $V_{o,e}$ and $V_{o,h}$ to vary within 25$\%$ of their \highlight{proposed} values, \high{and} \highlight{then we fitted Eq. \eqref{tauc} \high{to the data by varying} $e_{SPDEM}$ and $V_{o,e(h)}$ for each \highl{experimental} QD $D$ \highl{and $E_G(D)$}.} 

\section{Results \label{result}}

\begin{figure}
\includegraphics[scale=.7]{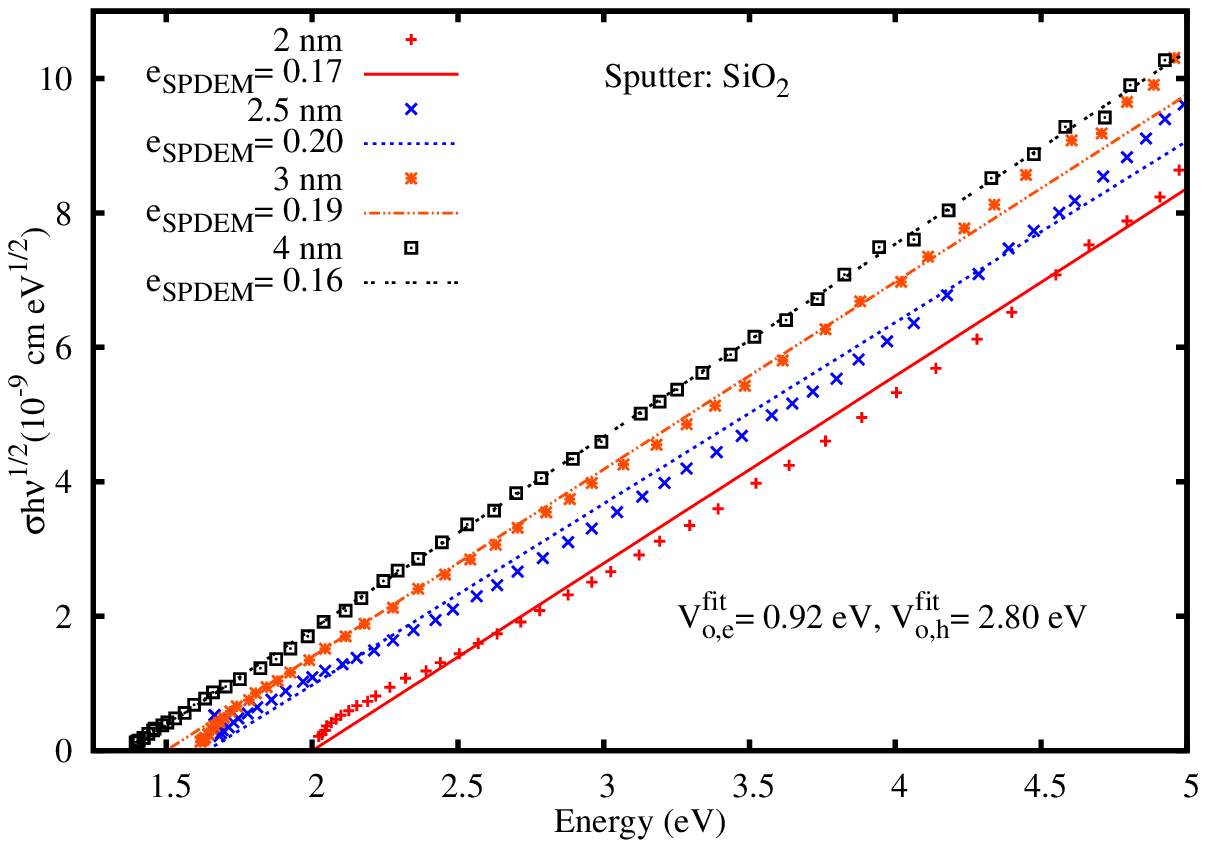}
\caption{Tauc plot of the absorption cross section with corresponding linear fit for sputtered Ge QDs from Ref. \onlinecite{Cosentino:2014_1}. The diameter for each sample is given in the legend, with the fitting parameter $e$ from Eq. \eqref{MDD}. The fitted interface energy from Eq. \eqref{SPDEM} is labelled in the plot.\label{sputt_abs}}
\end{figure}

\begin{figure}
\includegraphics[scale=.7]{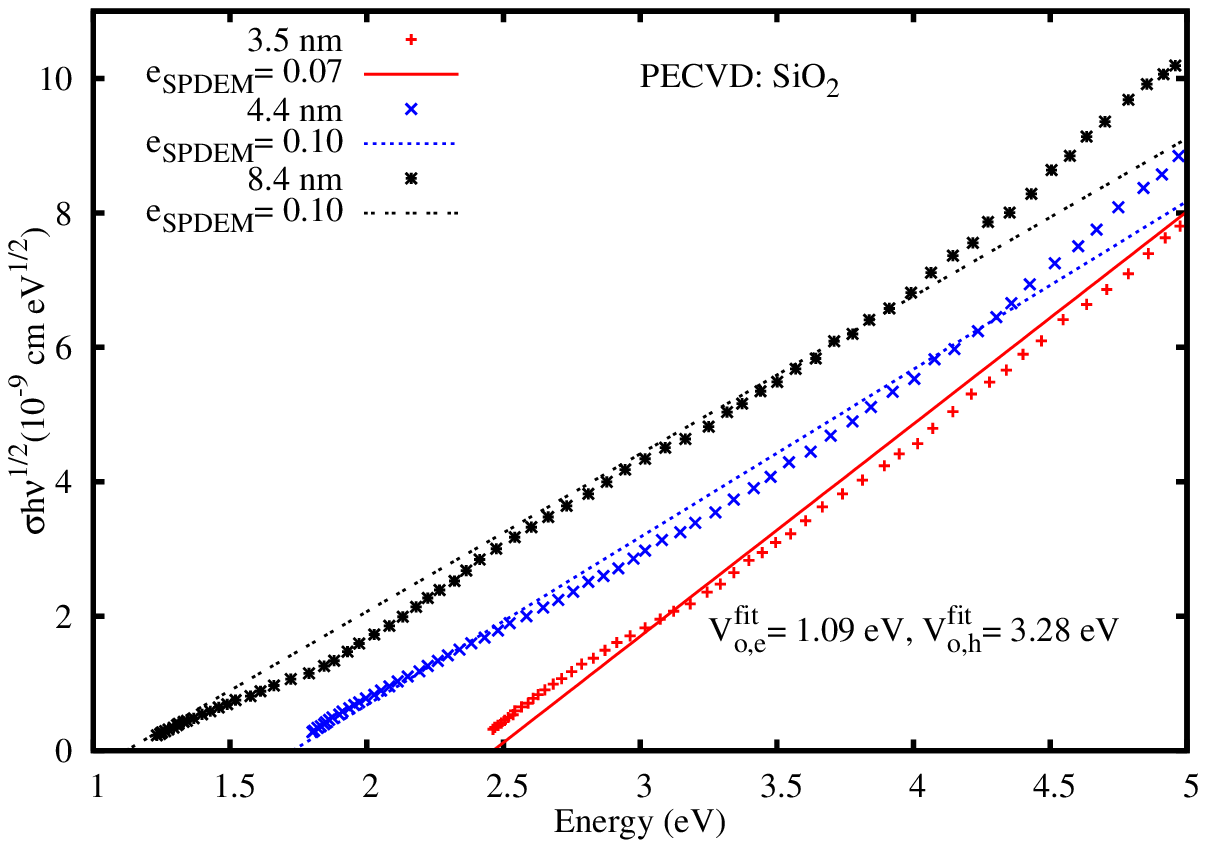}
\caption{Tauc plot of the absorption cross section with corresponding linear fit for PECVD Ge QDs from Ref. \onlinecite{Cosentino:2014}. The diameter for each sample is given in the legend, with the fitting parameter $e$ from Eq. \eqref{MDD}. The fitted interface energy from Eq. \eqref{SPDEM} is labelled in the plot.\label{PECVD_abs}}
\end{figure}

The results \markup{obtained from} fitting SPDEM-$\tilde{\mu}(D)$ \markup{to the sputtered and PECVD samples \markup{are} shown in Figs. \ref{sputt_abs} and \ref{PECVD_abs}, respectively}. \highl{The deviation from linearity in the experimental data at low $\sigma h\nu$ is due to Urbach tail regions, i.e. defect states near the band edge, which are common to amorphous structures as in this work \cite{Cosentino:2014_1}. In PECVD samples (Fig. \ref{PECVD_abs}) there is a deviation from linearity at high $\sigma h\nu$ due to a partial crystalline transition in less than 30$\%$ of the QDs after annealing at 800 $^o$C \cite{Cosentino:2014}. When the fit was constrained to the linear region, a less than 1 $\%$ difference was found in our final fitting parameters compared to fitting over the entire data set.} The \high{fitted} value for $e_{SPDEM}$ \markup{is} listed for each QD $D$ in Figs. \ref{sputt_abs} and \ref{PECVD_abs}, and the \high{resultant} interface potential \high{(also shown)}, \highlight{was found to be \high{notably} identical \high{(within 1$\%$)} for each QD $D$}. $e_{SPDEM}$ for each QD $D$ are the same within experimental error, therefore, in our formulation of $E_G(D)$ we use the average value, \markup{$\avg{e}$=0.181, and 0.090 for sputter, and PECVD, respectively}. \highlight{Likewise, $e_{EMA}$ varies only within experimental error.} The results of our \markup{analysis} are \markup{summarized} in Table \ref{parameters}. Table \ref{parameters} also lists the \highlight{results from the EMA fit and the values for} MBE Ge QDs.
\begin{table}
\caption{Renormalized mass parameter, $e$, for SPDEM (Eq. \eqref{SP_mu}) and EMA (Eq. \eqref{EM_mu}), and interface potential from Eq. \eqref{SPDEM} for sputter, PECVD, and MBE Ge QDs. \label{parameters}}
\begin{ruledtabular}
\begin{tabular}{c c c c c c} 
&\multicolumn{5}{c}{Sputter}\\
\cline{2-6}
{} & {2 nm} & {2.5 nm} & {3 nm} & {4 nm} & {$\avg{e}$}\\
\hline
{$e_{EMA}$} & {0.528} & {0.597} & {0.583} & {0.499} & {0.552} \\
{$e_{SPDEM}$ (nm$^{-1}$)} & {0.173} & {0.195} & {0.191} & {0.163} & {0.181} \\ 
&\multicolumn{5}{c}{$V^{fit}_{o,e}$=0.92 eV \quad $V^{fit}_{o,h}$=2.80 eV}\\
\cline{2-6}
&\multicolumn{5}{c}{PECVD}\\
\cline{2-6}
{} & {3.5 nm} & \multicolumn{2}{c}{4.4 nm} & {8.4 nm} & {$\avg{e}$} \\
\hline
{$e_{EMA}$} & {0.203} & \multicolumn{2}{c}{0.268} & {0.288} & {0.253} \\
{$e_{SPDEM}$ (nm$^{-1}$)} & {0.072} & \multicolumn{2}{c}{0.095} & {0.102} & {0.090} \\
&\multicolumn{5}{c}{$V^{fit}_{o,e}$=1.09 eV \quad $V^{fit}_{o,h}$=3.28 eV}\\
\cline{2-6}
&\multicolumn{5}{c}{MBE}\\
\cline{2-6}
{$e_{EMA}$} & \multicolumn{5}{c}{0.378} \\
{$e_{SPDEM}$ (nm$^{-1}$)} & \multicolumn{5}{c}{0.059} \\
&\multicolumn{5}{c}{$V_{o,e}$=0.20 eV \quad $V_{o,h}$=0.66 eV}
\end{tabular}
\end{ruledtabular}
\end{table}

\begin{figure}
\includegraphics[scale=.7]{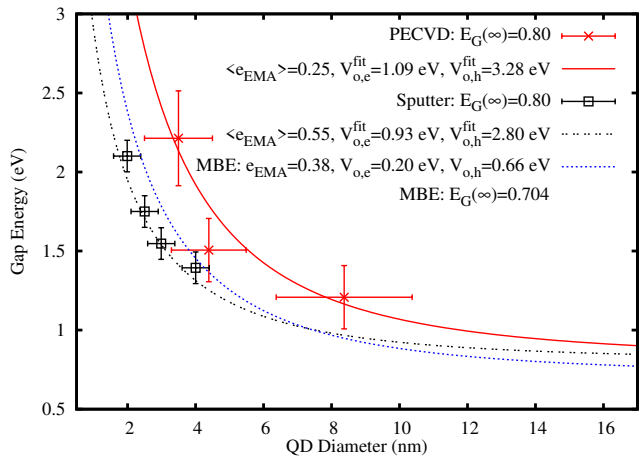}
\caption{Variation in the gap energy as a function of QD diameter for PECVD \cite{Cosentino:2014} and sputter \cite{Cosentino:2014_1} Ge QDs. \highlight{The theoretical curves are from Eq. \eqref{SPDEM} with the correction factor Eq. \eqref{MDD}. The MBE curve is taken from Ref. \onlinecite{Barbagiovanni:2013_3} for reference.} The bulk gap energy $E_G(\infty)$, interface energy, and $e$ is given for each sample \highlight{from Table \ref{parameters}}. The MBE \highlight{curve} uses the Ge/a-Si interface energy \cite{Barbagiovanni:2013_3}.
\label{QC_SPDEM}}
\end{figure}

In Fig. \ref{QC_SPDEM}, we plot the experimental data for PECVD \cite{Cosentino:2014} and sputter \cite{Cosentino:2014_1} Ge QDs along with the SPDEM-$\tilde{\mu}(D)$ curve. The interface potential, $e_{EMA}$, and $E_G(\infty)$ are listed in Fig. \ref{QC_SPDEM} for reference. In Sec. \ref{theory}, we mentioned that the SPDEM-$\tilde{\mu}(D)$ and EMA-$\mu(D)$ models agree with each other. The parameters $a$, $b$, and $c$ in Eqs. \eqref{SP_mu} and \eqref{EM_mu} were found by analysing MBE grown Ge QDs \cite{Barbagiovanni:2013_3}. There is good agreement between the experimental data and our calculation, noting that the error bars in the $D$ of the PECVD samples are on the order of 2 nm \cite{Cosentino:2014}.

The $E_G$ for the PECVD grown samples is higher than the sputter samples. The reason for the difference is understood within our models. From the SPDEM-$\tilde{\mu}(D)$ model, we find a larger interface potential in PECVD samples compared to the sputter samples. The larger interface potential is correlated with a larger reduction in $e_{EMA}$, which increases the confinement energy. MBE Ge QDs were fabricated with an a-Si matrix, which has a lower interface potential than both PECVD and sputter Ge QDs, see Fig. \ref{QC_SPDEM}. However, the $E_G$ is comparable to the other two samples with a value of $e_{EMA}$ in between the PECVD and sputter values. \highlight{It is difficult to make a \high{full comparison} between the sputter and PECVD samples versus the MBE samples, because the MBE samples were fabricated under very different conditions. However, we can consider a few essential differences.} In the case of the MBE QDs there are no O-defect states \cite{Hdiy:2008, Barbagiovanni:2013_3}, because of the unique fabrication method. O-defect states at the interface increase the spread of the wave-function, thus lowering the confinement energy \cite{Guerra:2011}. Furthermore, O interface states can pin the charge carriers, again lowering the confinement energy \cite{Barbagiovanni:2012, Barbagiovanni:2013}. \markup{The combination of these phenomena lowers the confinement \highlight{energy} in PECVD and sputter \highlight{samples, in agreement with theoretical predictions \cite{Guerra:2011}, and \high{it} becomes} comparable to that in MBE \highlight{samples} (see Fig. \ref{QC_SPDEM}).} \highlight{Therefore, we understand that the O interface states are the source of the difference between the sputter and PECVD samples versus the MBE samples.} \markup{Nonetheless}, we see that for a higher interface energy a larger reduction in the EM \markup{is obtained}.

\section{Discussion \label{disc}}

\begin{figure}
\includegraphics[scale=.75]{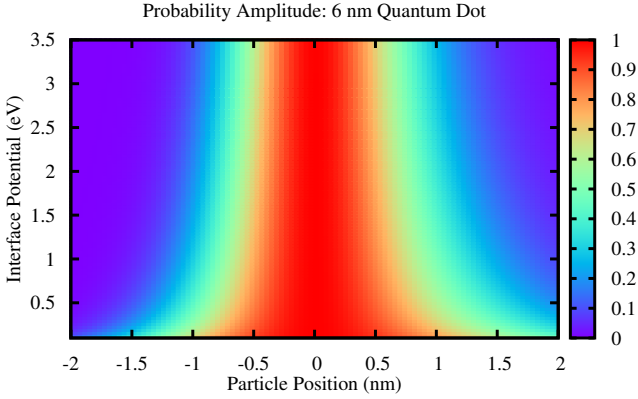}
\caption{Probability amplitude of Eq. \eqref{psi} taken from Ref. \onlinecite{Barbagiovanni:2013_1} for a 6 nm Ge QD as a function of particle position and interface potential. \markup{The} figure depicts increasing localization of the probability amplitude as a function \markup{of} increasing interface potential. Color online.\label{prob_amp_vo}}
\end{figure}

A closer examination of the results \markup{shown in} Table \ref{parameters} in the context of our models allows us to understand the nature of the confinement mechanism. As mentioned in Sec. \ref{theory}, the SPDEM theory models a modification in the EM due to the interface potential. In Ref. \onlinecite{Barbagiovanni:2013_1}, we derive the envelope function \markup{within} the SPDEM formalism:
\begin{equation}\label{psi}
\Psi_{i}(r_i)=\left(\frac{2}{\sigma_i^2\pi}\right)^{1/4}\exp\left(-\frac{1}{2\sigma_i^2\gamma_i^2}\ln^2(1+\gamma_i r_i)\right);
\end{equation}
where $\sigma\sim (V_o)^{-1/2}$ is the Gaussian strength. In Fig. \ref{prob_amp_vo}, we plot the probability amplitude of Eq. \eqref{psi} for a 6 nm QD as a function of the interface potential and particle position. As the interface potential increases, the probability amplitude is increasingly localized, which implies increasing de-localization in momentum space. \highlight{A Fourier transformation of Eq. \eqref{psi} demonstrated that the SPDEM envelope function exhibits \high{an} increased spread in momentum space compared to a Gaussian envelope function \cite{Barbagiovanni:2013_1}. The interface potential alters the SPDEM envelope function because of the introduction of $\gamma$, which modified the EM through the relation \cite{Barbagiovanni:2013_1}:}
\begin{equation}\label{EM_SPDEM}
m(r_i)=\frac{m_o^*}{\left(1+\gamma_i r_i\right)^2}.
\end{equation}

\begin{figure}
\includegraphics[scale=.7]{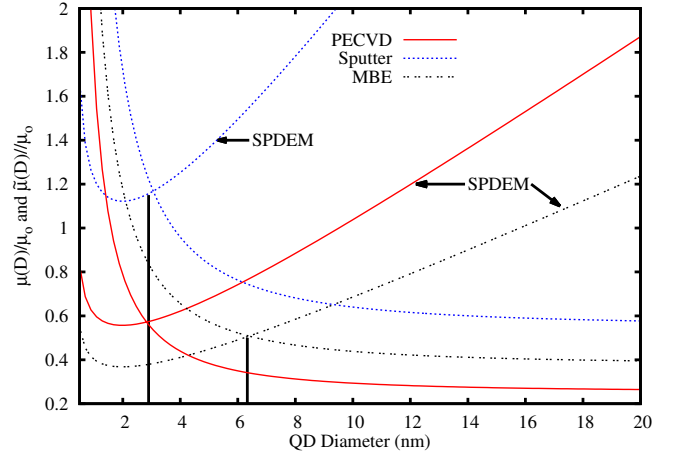}
\caption{Plot of Eqs. \eqref{SP_mu} and \eqref{EM_mu} normalized to the bulk reduced effective mass, $\mu_o$, for the different Ge QD samples as a function of QD diameter. The SPDEM-$\tilde{\mu}(D)$ (Eq. \eqref{SP_mu}) curves are indicated by an arrow. The vertical lines indicate the dimension where the behaviour changes from SPDEM to EMA.  \label{mass_ren}}
\end{figure}

Therefore, from $(e_{SPDEM})^{-1}$ we \markup{can} find the length scale \markup{at which} the EM begins to change, which is \markup{16.95}, 11.11, and 5.53 nm in \markup{MBE}, PECVD, and sputter \markup{samples}, respectively. At the stated dimensions the SPDEM energy begins to deviate from the experimental data and $e_{SPDEM}$ corrects for this behaviour. \markup{In Fig. \ref{mass_ren} we plot \highlight{$\tilde{\mu}(D)/\mu_o$ and} $\mu(D)/\mu_o$ from Eqs. \eqref{SP_mu} and \eqref{EM_mu}, respectively,} where the role of $e_{SPDEM}$ can be seen. Below diameters of around 16 and 11 nm the \highlight{MBE and PECVD} $\tilde{\mu}(D)/\mu_o$ curves are less than 1 indicating \highlight{that $\tilde{\mu}(D)$ increases} $E_G(D)$ \highlight{below these dimensions}. The difference between the \highlight{cross-over} dimension depicted in Fig. \ref{mass_ren} and the value \highlight{obtained} from $(e_{SPDEM})^{-1}$ is because of experimental error. We note that in Fig. \ref{mass_ren} the $\tilde{\mu}(D)/\mu_o$ curve for the sputter sample does not fall below 1. In Ref. \onlinecite{Barbagiovanni:2013_3}, we determined that above 6.2 nm the dispersion relation is dominated by the SPDEM-$\tilde{\mu}(D)$ relation. Therefore, the sputter $(e_{SPDEM})^{-1}$ value of 5.53 nm indicates that \highlight{the} SPDEM mechanism is suppressed in these samples. The different behaviour \highlight{found} for the sputter samples is a \highlight{consequence} of their interface quality (Secs. \ref{method} and \ref{result}). The lower interface potential does not create a strong confinement condition for the charge carriers, see Fig. \ref{prob_amp_vo}. Additionally, the sub-oxide interface states promote de-localization of the wave-function in real space \cite{Guerra:2011}.

\highlight {The values obtained for $(e_{SDPEM})^{-1} $ also agree with the results of Fig. \ref {QC_SPDEM}. First, we note that the value obtained for the MBE samples of $\sim $ 16 nm is on the order of the Bohr radius for Ge. This result is what we expect in these samples, because of the high quality interface. At 16 nm, the charge carriers in the MBE samples begin to experience the influence of the confinement potential and thus their energy is modified, as predicted by QC. While in the PECVD and sputter samples, the onset of QC is suppressed from the expected Bohr radius value.  Therefore, the PECVD and sputter samples experience the effects of QC over a smaller diameter range compared to MBE, \high{and} thus their $E_G$ shift is not as great.}

The relationship between pure QC effects and a modified EM are understood by comparing the SPDEM-$\tilde{\mu}(D)$ results with respect to the EMA-$\mu(D)$ results. In a previous work on Ge QDs buried in a-Si \cite{Barbagiovanni:2013_3} we found that above $\approx$6.2 nm the $E_G\sim D^{-1}$, due to a modification in the EM, while below $\approx$4.6 nm the $E_G\sim D^{-2}$, due to pure QC effects. These parameters come from the term $\left(1+\left (aD^2+bD+c\right)^{-1}\right)$ in Eqs. \eqref{SP_mu} and \eqref{EM_mu}, and \textit{represent a universal dimensional dependence}. The parameter $e$ represents a correction to these ideal length scales due to additional effects not included in the theory, which in our analysis include the O-interface states. If we consider the ratio $e_{EMA}/e_{SPDEM}$, then we obtain 3.05 and 2.81 nm for sputter and PECVD samples, respectively. \markup{This length scale signifies when the confinement behaviour changes from the SPDEM to the EMA model and is depicted in Fig. \ref{mass_ren} by the vertical line at 2.9 nm.} Within experimental error, we thus find that pure QC effects dominate at $\approx$ 2.9 nm, which is suppressed from the ideal case (between 6.2 and 4.6 nm) due to the interface states. On the other hand, the same ratio in MBE samples gives 6.4 nm (see vertical line in Fig. \ref{mass_ren}), which is in agreement with the ideal value (6.2 nm), because there are no O interface states in this sample. \highlight{Finally, \high{this is} in agreement with previous results \high{obtained} in Ref. \onlinecite{Barbagiovanni:2013_3}, \high{where} the MBE value of 6 nm was found to be the dimension that produces the most efficient luminescence. From  our work here we understand this result on the basis that this is the dimension \high{at which} the behaviour changes from SPDEM to EMA and thus carriers are strongly confined.}

\highlight{We illustrate collectively \high{all of the} phenomena described in this manuscript in Fig. \ref{interface_em}. At the top of the figure we recall the dimensional dependence of the SPDEM-$\tilde{\mu}(D)$ and EMA-$\mu(D)$ models. Fig \ref{interface_em} is divided \high{into} two parts. In the `pink' region of the figure above $\sim$ 6 nm, the dominant model is SPDEM-$\tilde{\mu}(D)$, which models a change in $E_G$ in conjunction with a change in the EM. The `grey' region below $\sim$ 4 nm is modelled according to the EMA-$\mu(D)$ where pure QC effects dominate. The `pink' and `grey' regions were determined from the results of Ref. \onlinecite{Barbagiovanni:2013_3} and come from the universal dimensional parameters $a$, and $b$ in the term $\left(1+\left (aD^2+bD+c\right)^{-1}\right)$.
\begin{figure*}
\includegraphics[width=.9\textwidth]{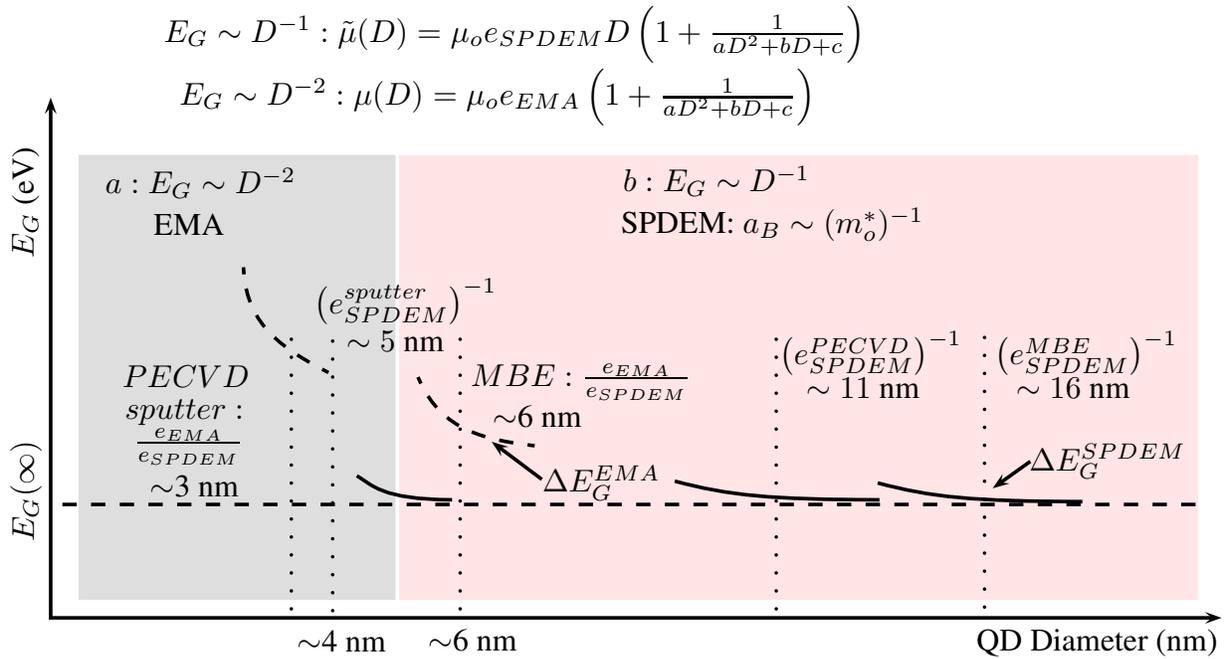}
\caption{Schematic illustration of the confinement mechanism. The dimensional behaviour of the SPDEM-$\tilde{\mu}(D)$ and EMA-$\mu(D)$ is noted at the top of the figure. In the `pink' region above $\sim$ 6 nm the dominate model is SPDEM, while in the `grey' region below $\sim$ 4 nm EMA is the dominant model, see Ref. \onlinecite{Barbagiovanni:2013_3}. The bulk $E_G$ is represented by the straight dashed line. At $\sim$ 16 , 11, and 5 nm the MBE, PECVD, and sputter samples, respectively, experience a change in the EM and hence $E_G$, according to the SPDEM model. The change in energy due to SPDEM is represented by the solid curved line in the three regions for the respective samples (not drawn to scale). At $\sim$ 6, and 3 nm the MBE, and PECVD/ sputter samples, respectively, change behaviour from SPDEM to EMA. The change in energy due to EMA is represented by the dashed curved line in the two regions for the respective samples (not drawn to scale). The lower dimensions in PECVD and sputter samples compared to the predicted pink and grey regions is a result of the O interface. Color online.\label{interface_em}}
\end{figure*}

From the parameter $(e_{SPDEM})^{-1}$, we determined the onset of the SPDEM behaviour at $\sim$ 16, 11, and 5 nm in the MBE, PECVD, and sputter samples, respectively. The change in energy due to SPDEM is represented by the solid curved line in Fig. \ref{interface_em} for the three samples (not drawn to scale). The MBE \high{sample's} value is in agreement with the Bohr radius for Ge and indicates when the charge carriers first `feel' the confinement potential. In the case of PECVD and sputter \high{samples} this value is suppressed due to the O states at the interface, which cause the wave-function to leak into the matrix and thus lowers the confinement energy, as described above. \high{The} PECVD \high{case} has a higher value than \high{the} sputter \high{case} due to the higher confinement potential and thus experience a larger change in $E_G$, due to an increased reduction in the EM.

The transition \high{point} from SPDEM to EMA is on the order of 5 nm according to the transition from the `pink' to the `grey' region in Fig. \ref{interface_em}. From the ratio of $e_{EMA}$ to $e_{SPDEM}$ we determined this transition for each sample. The change in energy due to EMA is represented by the dashed curved line in Fig. \ref{interface_em} for the three samples (not drawn to scale). In the case of MBE \high{sample}, the transition is in agreement with the calculated value at 6 nm, because of the lack of O at the interface. While in PECVD and sputter \high{samples} they both change behaviour at around 3 nm. The suppression here again is due to O interface states, as described above.}

\section{Conclusion \label{conc}}

Using a previously developed theoretical formalism \highli{as an analytical tool} we \high{have been} able to understand the relationship between the EM and the interface potential. We found that when the interface potential was large there exists a strong reduction in the EM. This condition was clearly demonstrated when comparing PECVD and sputter samples, where \high{the} PECVD \high{case} has a larger interface potential. However, if the sample contains O states at the interface, then the confinement energy was reduced due to these states. This result was noted by comparing MBE samples with no O interface states to the PECVD and sputter samples. We determined the length scale where the samples begin to experience a modification in $E_G$ due to a change in the EM, which was $\sim$ 16, 11, and 5 nm for MBE, PECVD, and sputter \high{samples}, respectively. The lower values in PECVD and sputter \high{samples} compared to MBE \high{samples} were understood in terms of the O interface states. We found the length scale where the QC mechanism changes from SPDEM to EMA, which was 6, and 3 nm in MBE and PECVD/sputter \high{samples}, respectively. All of these results are summarised \high{graphically} in Fig. \ref{interface_em}. These results indicate an additional \high{carrier} confinement mechanism in NSs, which can be exploited for device fabrication through interface engineering. \high{From} this work, we \high{have} found a dynamic relationship between the magnitude of the interface potential and the EM with respect to the chemical composition of the interface.

\acknowledgements 
We would like to thank N. L. Rowell from NRC-Ottawa; I. Berbezier, G. Amiard, L. Favre, and A. Ronda from UMR CNRS-Marseille; and M. Faustini, and D. Grosso from UMR-7574 UPMC-CNRS-Paris for their valuable experimental contributions and insights with respect to MBE grown Ge QDs. R.N.C.F. acknowledges funding from CNPq (Brazilian agency). S.C., A.T., and S.M. acknowledge MIUR projects, and ENERGETIC (PON00355$\textunderscore$3391233). E.G.B. and S.M. acknowledge PLAST$\textunderscore$ICs (PON02$\textunderscore$00355$\textunderscore$3416798).

%\bibliography{refr}

%merlin.mbs aipnum4-1.bst 2010-07-25 4.21a (PWD, AO, DPC) hacked
%Control: key (0)
%Control: author (8) initials jnrlst
%Control: editor formatted (1) identically to author
%Control: production of article title (-1) disabled
%Control: page (0) single
%Control: year (1) truncated
%Control: production of eprint (0) enabled
%

\end{document}